\documentclass[onecolumn,authoryear]{els-mrw} 

\usepackage{amsmath,amssymb,amsfonts,amsthm,makeidx,graphicx}
\usepackage{txfonts}
\usepackage{helvet}

\newcommand{\msun}{\mbox{$M_{\odot}$}}
\newcommand{\Msun}{\mbox{$M_{\odot}$}}

\newcommand{\mdot}{\mbox{$\dot{M}$}}

\newcommand{\msunyr}{\mbox{$M_{\odot} {\rm yr}^{-1}$}}

\newcommand{\ha}{H$\alpha$}

\newcommand{\apj}{ApJ}
\newcommand{\apjl}{ApJL}
\newcommand{\aap}{A\&A}
\newcommand{\mnras}{MNRAS}
\newcommand{\araa}{ARA\&A}
\newcommand{\aapr}{A\&ARev}
\newcommand{\nat}{Nature}

\begin{document}

\chapter{Stellar Winds}\label{chap1}

\author[1]{Jorick S. Vink}%

\address[1]{\orgname{Armagh Observatory and Planetarium}, \orgdiv{Division or Department}, \orgaddress{College Hill, BT61 9DG Armagh}}
%\address[2]{\orgname{Name of Institute}, \orgdiv{Division or Department}, \orgaddress{Address of Institute}}

\articletag{Chapter Article tagline: update of previous edition,, reprint..}

\maketitle

\begin{glossary}[Glossary]

\term{Stellar winds} Outflows from all types of stars, driven by a plethora of physical mechanisms; sometimes modest in affecting its own evolution, sometimes highly influential.

\term{Solar Wind} Stellar wind driven by gas pressure; relatively modest in terms of its absolute mass-loss rate.

\term{CAK theory} Radiation-driven wind theory for massive OB stars; relevant for stellar evolution of massive stars.

\term{P Cygni} Spectral line involving nearly symmetric line emission plus blue-shifted absorption. Named after the LBV star P\,Cygni. 

\term{Sobolev approximation} Simplified radiative transfer when the velocity gradient is large, and local physical conditions remain constant.

\term{Multiple scattering} The ability of photons to transfer energy and momentum several times.

\term{Clumping} Wind structures that appear on small scales. If the photons leak through the clumps, one needs to consider porosity as well.

\term{Wolf-Rayet stars} classical WRs come in two main types WN and WC. Nitrogen-rich WN stars have enhanced nitrogen (N) from the CNO cycle; the more evolved WC stars show carbon (C) at the surface.

\term{WNh stars} Main sequence N-rich WRs that still contain hydrogen (H). The less dense-winded Of/WN "slash" stars are also included.

\term{Transition mass-loss rate} Point where an optically-thin CAK wind kinks upwards into an optically thick WNh wind. 

\term{Transformed mass-loss rate} A useful 
concept for setting up atmospheric wind model grids.

\term{Superwind} continuously increasing wind mass-loss strength during later stellar evolution.

\term{Super-Eddington wind} A radiation-driven wind that might occur when stellar atmospheres are formally above the Eddington limit.

\end{glossary}

\begin{glossary}[Nomenclature]
\begin{tabular}{@{}lp{34pc}@{}}
CAK & Castor, Abbott \& Klein (1975)\\
VMS & Very Massive Stars\\
WR  & Wolf-Rayet star\\
LBV & Luminous Blue Variable\\
AGB & Asymptotic Giant Branch \\
RSG & Red Supergiant\\
NLTE & Non Local Thermodynamic Equilibrium\\
CMF & Co-moving Frame\\
\end{tabular}
\end{glossary}

\begin{abstract}[Abstract]
Stellar winds form an integral part of astronomy. The solar wind affects Earth's magnetosphere, while the winds of hot massive stars are highly relevant for galactic feedback through their mechanical wind energy. In different parts of the stellar HR diagram different forces dominate. On the hot side of the HRD radiative forces on ionised gas particles are active, while on the cool side molecular and dust opacities take over. Moreover, due to the convective envelopes, alternative physical ingredients may start to dominate. I will describe the basic equation of motion and give a few examples, mostly focusing on the winds from massive stars. Here mass-loss rates significantly affect the stellar evolution all the way to core collapse as a supernova and/or black hole formation event.
\end{abstract}

\begin{BoxTypeA}[chap1:box1]{Key Points}

\begin{itemize}
%\end{BoxTypeA}    
%\section*{Box 1 hd}
%Stated another way, this concept requires that the mass into and out of an infinitesimal box must be equal to the change of mass in the box. Such a volume is sketched,
%\subsection*{Box 2 hd}
%The symbols $\delta x$, $\delta y$, and $\delta z$ represent the
\item The physical mechanism for the winds of low-mass stars, such as cool stars those on the AGB -- that are thought to undergo a superwind -- are understood to be driven by radiation pressure on dust.

\item The situation for massive cool stars, such as RSGs is far less clear, but a superwind might be relevant in this part of the stellar HR diagram as well.

\item Hot-star winds are thought to be driven by radiation pressure on gas, but the absolute mass-loss rates are uncertain due to wind clumping.

\item Uncertainties are largest below $\simeq 20 \msun$, substantial (by a factor 3 or so) for canonical massive stars in the 20-60 \msun\ range, and most accurate (about 30\%) for VMS at the 
transition mass-loss point. Here, the effects of winds on stellar evolution are the most dramatic.

\end{itemize}

\end{BoxTypeA}

\section{Introducing stellar wind observations
\label{chap1:sec1}
}

While we directly notice the effects of the solar wind on our Earth's atmosphere, how do we know that other types of stars lose mass in winds as well?
There are several ways in which stellar winds of low-mass and high-mass stars are probed, and I refer to \cite{vidotto} for overviews on winds from stars as well as exo-planets. 
For stars like the Sun, the mass-loss rate is of order $10^{-14} \msunyr$ (see Table 1) and even for the total 10 billion year lifetime of the Sun, the total mass lost during solar evolution is a mere $\sim$ 0.0001\msun. 
However, for massive O-type stars in the range 20-100\msun\ mass-loss rates are of the order of $10^{-7} - 10^{-4} \msunyr$ and even considering the shorter evolutionary lifespan of millions of years, high-mass stars lose quantities that are sufficiently high to affect their own stellar evolution considerably. 
Very massive stars (VMS) above 100\Msun\ at the high end of the massive star range are even expected to evaporate almost entirely. 
For stellar evolutionary calculations a key interest is how the mass-loss rates depends on the stellar parameters, such as mass $M$, associated luminosity $L$, effective temperature $T_{\rm eff}$, and metallicity $Z$, i.e. $\dot{M} = f(Z,L,M,T_{\rm eff})$.

The division between solar-type and high-mass stars does not imply that the evolution of low-mass stars is not affected by winds at all. 
In fact, during the core helium (He) and later evolutionary phases, especially when they approach the asymptotic giant branch (AGB) in the cool part of the stellar HR diagram, the mass-loss rates become so strong, of order $10^{-6} \msunyr$ \citep{hofner} that AGB winds remove the outer stellar layers, producing hot white dwarfs. This rapidly increasing mass-loss rate (reaching $10^{-4} \msunyr$) towards the top end of the AGB luminosity function is usually referred to as the superwind phase.

\begin{table}
\tabcolsep7.5pt
\caption{Typical wind parameters for different stellar Types. The numbers should only be taken as typical.}
\begin{center}
\begin{tabular}{@{}l|c|c|c|c|l}
\hline
Type   & $T_{\rm eff}$ & $M$           & $v_{\infty}$ & $\dot{M}$  \\
       &  (kK)         & ($M_{\odot}$) & (km/s)       & ($\msunyr$) \\
\hline
Sun      & 6           &  1            & $\sim$500    & $10^{-14}$\\
O        & 30-45       &  20-60        & 2000-3500    & $10^{-7}-10^{-5}$ \\
\hline
Of/WN    & 35          & 80            & 2000         & $10^{-5}$\\
\hline
WNh      & 35-50       &  80-300       & 1500-3000    & $10^{-5}$-$10^{-4}$\\
BSG      & 15-25       &  15-30        & 500-1500     & $10^{-7}-10^{-5}$ \\
YSG      &  5-10       &  10-25        & 50-200       & $10^{-6}-10^{-4}$ \\
RSG      &  3-4       &   10-25        & $\sim$30        & $10^{-7}-10^{-3}$ \\
AGB      & 3-4 & 1-8 & $\sim$10 & $10^{-7}-10^{-4}$\\
LBV low-L & 10-15     &   15-25        & 100-200          & $10^{-5}$         \\
LBV high-L & 10-30     &   40-        & 200-500          & $10^{-4}-10^{-3}$  \\
cWR        & 90-200     &  10-30        & 1500-6000    & $10^{-5} - 10^{-4}$ \\
Stripped He  & 50-80   &  1-5          & 1000         & $10^{-8}$          \\
\hline
\end{tabular}
\end{center}
\end{table}

One of the key observational probes of such cool star mass loss is via their circumstellar dust emitted in the infrared. 
Similarly, massive red supergiants (RSGs) up to $\simeq$ 25\msun\ emit infrared radiation from dusty winds, but it is not yet established if this dust plays a key role in the wind driving, or that RSG winds are driven by another mechanism and that this outflow-generated material just happens to condense into the dust we observe \citep{decin}.
Massive stars on the main sequence are generally considered to be too hot to produce dust, and optical spectral line diagnostics is the prime modus operandum\footnote{There are alternative techniques from long-wavelength free-free emission in the sub-mm and radio range \citep{WB75,PF75}, as well as infrared line spectroscopy \citep{Najarro11}}. 

The mass-loss rates is generally expressed in terms of wind density $\rho$ and velocity $v$, as:

\begin{equation}
\dot{M}~=~4 \pi r^2 \rho(r) v(r)
\label{eq_mdot}
\end{equation}
As most diagnostics do not directly probe the mass-loss rate $\dot{M}$ but rather the wind density $\rho$, knowledge about the wind velocity is generally required from spectral kinematic information. This holds true for both continuum gas and dust techniques, as well as the optical line techniques discussed in the following.

Hot stars with dense winds, such as O-type supergiants, generally show H$\alpha$ emission lines resulting from hydrogen (H) recombination (see the middle panel (b) in Fig.\,1. The (negative) equivalent width (EW) of such an emission line -- as opposed to the positive EW from photospheric absorption (panel a) lines -- probes the density of the stellar wind as recombination scales with $\rho^2$. In principle the width of the line provides information on the wind velocity, but as the \ha\ line is formed deep inside the wind, where the wind has not yet reached its terminal velocity, this would underestimate $v_{\infty}$.
A more accurate determination of the terminal velocity is given by the blue edge of P\,Cygni lines, depicted in the right-hand side (panel c) of Fig.\,1. 
These resonance lines are formed by line scattering and are predominately located in the ultra-violet (UV) part of the electromagnetic spectrum. 

\begin{figure}
  \includegraphics[width=\columnwidth]{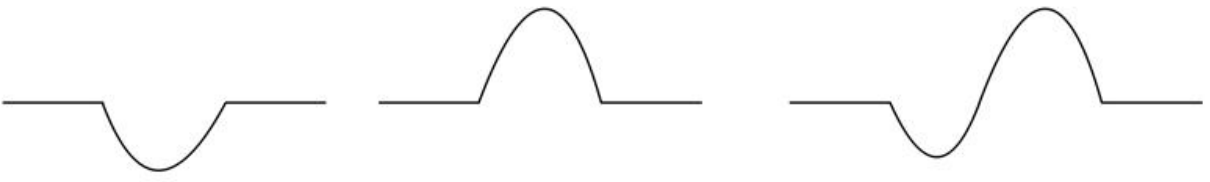}
  \caption{Flux versus wavelength. (a) A typical stellar absorption line; (b) a recombination emission line; (c) a P Cygni resonance line including both blue-shifted absorption and red-shifted emission.}
  \label{f_ethan}
\end{figure}

Returning to the determination of the mass-loss rate, the \ha\ emission lines provide the density, while the UV lines offer velocity information. This means that for stationary smooth spherical winds, the mass-loss rates can simply be determined from Eq.\,\ref{eq_mdot}.
We will see later that hot-star winds are not smooth but structured and that we require multi-line UV and optical diagnostics \citep[e.g.][]{bouret12} to measure the mass-loss rates from clumped winds.

\section{Stellar Wind Theory}

Stellar winds are outflows of plasma that have managed to overcome the gravitational pull from the star. While stellar winds are oftentimes intrinsically time dependent, meaningful models for stationary outflows have been developed over the past six decades, ever since the landmark paper by Eugene Parker in 1958 in which a modern model for solar and stellar winds was developed. For the solar wind the relevant force is the gas pressure (where the issue of what heats the hot corona is an additional problem). The equation of motion relates the inertia term ($v dv/dr$) to the gravity ($\frac{GM}{r^2}$) and additional accelerations from gas pressure ($p$) and other forces:

\begin{equation}
v~(\frac{dv}{dr})~=~-\frac{GM}{r^2}~-~\frac{1}{\rho} \frac{dp}{dr}~+~a(r)
\label{eq_eom}
\end{equation}
but where $a(r)$ can be an representative acceleration due to any additional force, such as radiation pressure (see below), rotation, magnetic Lorentz forces, etc. (see \cite{LC99} for more details on both force and energy requirements). The key point is that if one were to add an additional force below the sonic point of the wind, this will increase the star's mass-loss rates, while adding a force in the supersonic part of the wind will merely increase the object's wind velocity.

For hot OB-type stars the key addition to the equation of motion is the radiative force on line opacity \citep{CAK} or CAK. If we now ignore the gas pressure term (which is much smaller than the radiation pressure term above the sonic point of the wind) we find an equation of motion that reads:

\begin{equation}
v~(\frac{dv}{dr})~=~-\frac{GM}{r^2}~+~C~(\frac{dv}{dr})^{\alpha}_{\rm Sob}
\label{eq_eom_rad}
\end{equation}
Where $C$ is a constant. By equating the velocity gradient in the inertia term on the left-hand side of Eq.\,\ref{eq_eom} to the Sobolev\footnote{The Sobolev approximation is used to simplify radiative transfer when the velocity gradient $dv/dr$ is large. i.e. when conditions over the Sobolev length $v_{\rm th}/(dv/dr)$ – in terms of the thermal line width ($v_{\rm th}$) – hardly change.} velocity gradient on the right-hand side, CAK could 
solve the equation of motion analytically, and predict the wind velocity to behave as a "beta" law:

\begin{equation}
v(r)~=~v_{\infty}~(1 - \frac{R_{\ast}}{r})^{\beta} 
\end{equation}
\label{eq_vel}
with $\beta = 0.5$ and -- after some algebra -- a mass-loss rate $\mdot$ that depends on the stellar mass $M$, luminosity $L$, and the Eddington factor $\Gamma$, as  
\begin{equation}
\dot{M}~\propto~(kL)^{1/\alpha} M(1~-~\Gamma)^{1~-~1/\alpha}
\end{equation}
Where $k$ is a CAK parameter that depends on line strength, and where the Eddington factor $\Gamma$ is given as,  
\begin{equation}
	\Gamma  = \frac{\kappa}{4\pi c G}\frac{L}{M}
\label{eq_gamma}
\end{equation}
CAK-based models have been substantially modified using more complete atomic data, such as complex iron (Fe) ions. Moreover, the simplified point-source approximation was dropped resulting in larger $\beta$ values, of order unity. The terminal velocities were predicted to be a few times the escape speed, corresponding to the correct ballpark figures of 2000-3000 km/s \citep{pauldrach86}. I refer to \cite{puls08} and \cite{owocki15} for more extensive discussion of spectral line driving in CAK-based theories, as well as their associated instabilities, such as the "line-deshadowing instability" (LDI).

\section{Beyond CAK O star models}

Instead of parameterising the radiative acceleration using CAK force multipliers, the current state-of-the art -- in terms of hydrodynamic solutions approaches to the velocity stratification -- 
relies on either (i) the Lambert W function \citep{MV08} in which the radiative acceleration is determined from Monte Carlo methods \citep{LA93,vink2001}, or (ii) on the direct integration of the radiative acceleration in the co-moving frame (CMF):

\begin{equation}
 \label{eq:arad}
  a_{\rm rad} = \frac{1}{c} \int \kappa_\nu F_\nu {\rm d}\nu,
\end{equation}
where $\kappa_{\nu}$ is the frequency-dependent opacity and $F_{\nu}$ gives the flux. Full integrations of this equation have been undertaken by \cite{GH05}, \cite{sundqvist19} and \cite{krticka17}. 
The Sobolev approximation is no longer needed, and as long as the detailed atmospheric physics allows it, the CMF models are equally relevant for optically thin and thick winds, including the winds of Wolf-Rayet (WR) stars \citep{sander20}.

\subsection{Optically thick winds}  

Classical WR stars have winds with mass-loss rates typically a factor ten higher than canonical optically-thin O-star winds of the same luminosity.
Optically thick WR winds are not easily explained by the optically-thin CAK theory that entertains a velocity gradient $dv/dr$. 
In fact, for the optically thick WR winds $dv/dr$ is close to 0 \citep{GH05}. 
Empirical wind efficiency values of $\eta$ 

\begin{equation}
\eta = \frac{\mdot v_{\infty}}{L/c}
\end{equation}
are
typically in the range of 1-5, i.e. well above the single-scattering limit \citep{Crow07}.

\begin{figure}
  \includegraphics[width=0.9\columnwidth]{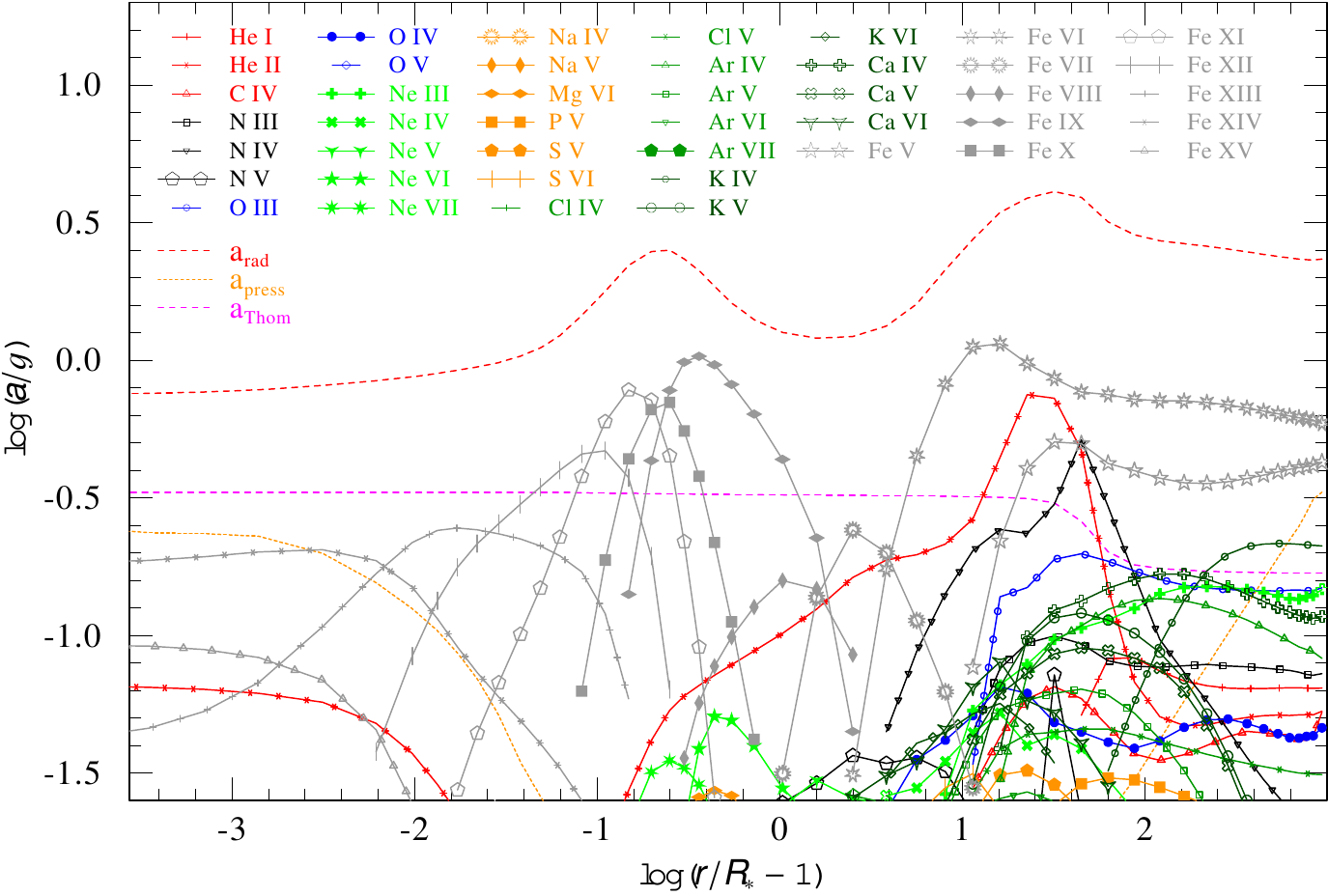}
  \caption{Contributions of the main driving ions to $a$ (in function of Newtonian gravity $g$) of a hydro-dynamically consistent N-type WR model through the atmosphere \citep{sander20}. The key contributions are indicated with different colours and symbols. The total radiative acceleration, the Thomson acceleration from free electrons, and the contribution from gas and turbulent pressure are also shown.}
  \label{f_leadions}
\end{figure}

As the degree of ionisation decreases radially outwards, photons may interact with opacity from a wide range of Fe and other relevant ions on their way through the atmosphere (see Fig.\,\ref{f_leadions}). Gaps 
between the spectral lines are ``filled in'' resulting from a layered ionisation structure \citep{LA93,Gayley95}. 
The initiation of classical WR outflows relies on
the condition that the winds are optically thick at the sonic point, and that the line acceleration due
to the high opacity Fe peak is able to counter-balance gravity, driving an optically thick wind \citep{NL02}. 

\begin{figure}
  \includegraphics[width=0.75\columnwidth]{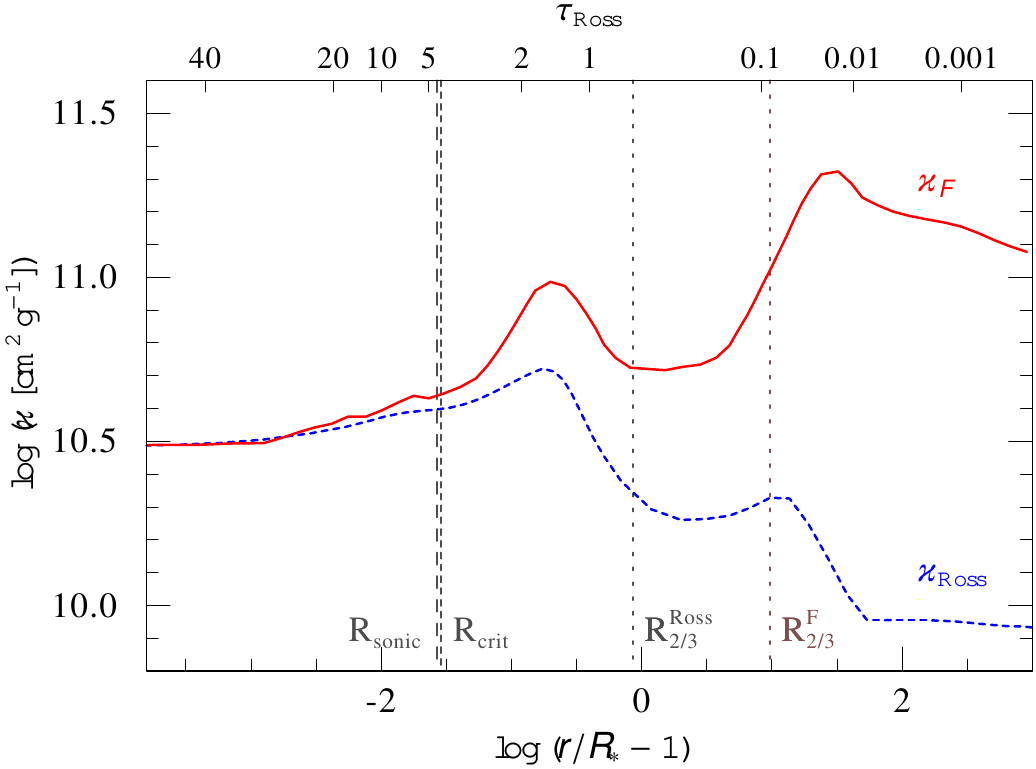}
  \caption{Comparison of the flux-weighted mean opacity (red, solid) to the Rosseland mean opacity (blue, dashed) in a model atmosphere characteristic of a classical WR star. Note that the opacities are equal to one another around the sonic/critical point, but that they diverge strongly -- by an order of magnitude -- in the outer wind. The two iron bumps discussed by \cite{NL02} are clearly present. See \cite{sander20} for more details.}
  \label{f_checkopaross}
\end{figure}

The key point of these optically thick wind analyses  
is that due to their huge mass-loss rates, the atmospheres become so dense that the wind sonic point is reached 
at very high flux-mean optical depth, and the radiation can be treated in the diffusion approximation. 
In this situation, the equation for the radiative acceleration can be simplified as:

\begin{equation}
 \label{eq:arad}
  a_{\rm rad} = \frac{1}{c} \int \kappa_\nu F_\nu {\rm d}\nu
  \simeq \kappa_{\rm Ross}\frac{L}{4\pi r^2 c},
\end{equation}
where $\kappa_{\rm Ross}$ is the Rosseland mean opacity (obtained from OPAL opacity tables). 
The $\Gamma$ limit with respect to the Rosseland mean opacity is crossed at the sonic point.
\cite{NL02} showed that the condition that $\kappa_{\rm Ross}$ needs to increase
outward with decreasing density could be fulfilled at the hot edges of iron opacity peaks, a ``cool'' bump at $\sim$ 70\,kK and a ``hot'' bump 
at 160\,kK, which can be identified as the 2 bumps in Fig.\,\ref{f_checkopaross}.  

Rather than considering $a_{\rm rad}$, it could be considered more practical to relate it back through the Eddington ratio
\begin{equation}
  \label{eq:gammarad}
	\Gamma_{\rm rad}(r) = \frac{a_{\rm rad}(r)}{g_{\rm grav}(r)} = \frac{\kappa_{F}(r)}{4\pi c G}\frac{L}{M}
\end{equation}
where $g_{grav} = - \frac{GM}{r^2}$ and the radial dependency of $\Gamma_{\rm rad}$ is due to the flux-weighted mean opacity $\kappa_{F}(r)$. 
In the deepest atmospheric layers the diffusion approximation is valid, and $\kappa_{F}$ approaches $\kappa_{\rm Ross}$ as depicted in Fig.\,\ref{f_checkopaross}. Note that differences between these two opacities become most notable further out in the wind, where they diverge by an order of magnitude. Hence, the modelling of an appropriate wind stratification involves an accurate calculation of $\kappa_{F}(r)$ through the entire atmosphere, from the deep optically thick layers to the outer optically thin regions.

\subsection{Transition mass-loss rate}
\label{sec_trans}

Until today, most stellar evolution models employ the \cite{vink2001} OB star prescription for main-sequence stars in terms of $\dot{M} = f(Z,L,M,T_{\rm eff})$. But there are still many uncertainties in the quantitative mass-loss rates of massive stars. 
Reasons for this vary from the usage of the Sobolev approximation to the role of wind clumping. 

In order to calibrate absolute mass-loss rates, \cite{VG12} developed the concept of the mass-loss transition point by solving the equation of motion (Eq.\,\ref{eq_eom}) in integral form. 
Employing the mass-continuity equation 
and the wind optical depth $\tau = \int_{r_s}^\infty \kappa_{\rm F}\rho\, {\rm d}r$, they derived:

\begin{equation}
\frac{\dot{M}}{L/c} dv = \kappa_{\rm F} \rho \frac{\Gamma - 1}{\Gamma} dr = \frac{\Gamma-1}{\Gamma} d\tau.
\label{eq_gaga}
\end{equation}
As they could assume that $\Gamma$ is
much larger than unity for O-stars in the supersonic region, the factor
$\frac{\Gamma-1}{\Gamma}$ would become approximately one\footnote{In reality this is not strictly true but needs a correction factor which was determined to be $0.6$ \citep{VG12}.}. Moreover, they could also show that:

\begin{equation}
\eta = \frac{\dot{M} v_{\infty}}{L/c} = \tau 
\label{eq_eta}
\end{equation}

and that one could simply employ the key condition $\eta = \tau = 1$ 
exactly at the transition
from optically thin O-star winds to optically-thick WR winds. 

\begin{figure}
  \includegraphics[width=0.75\columnwidth]{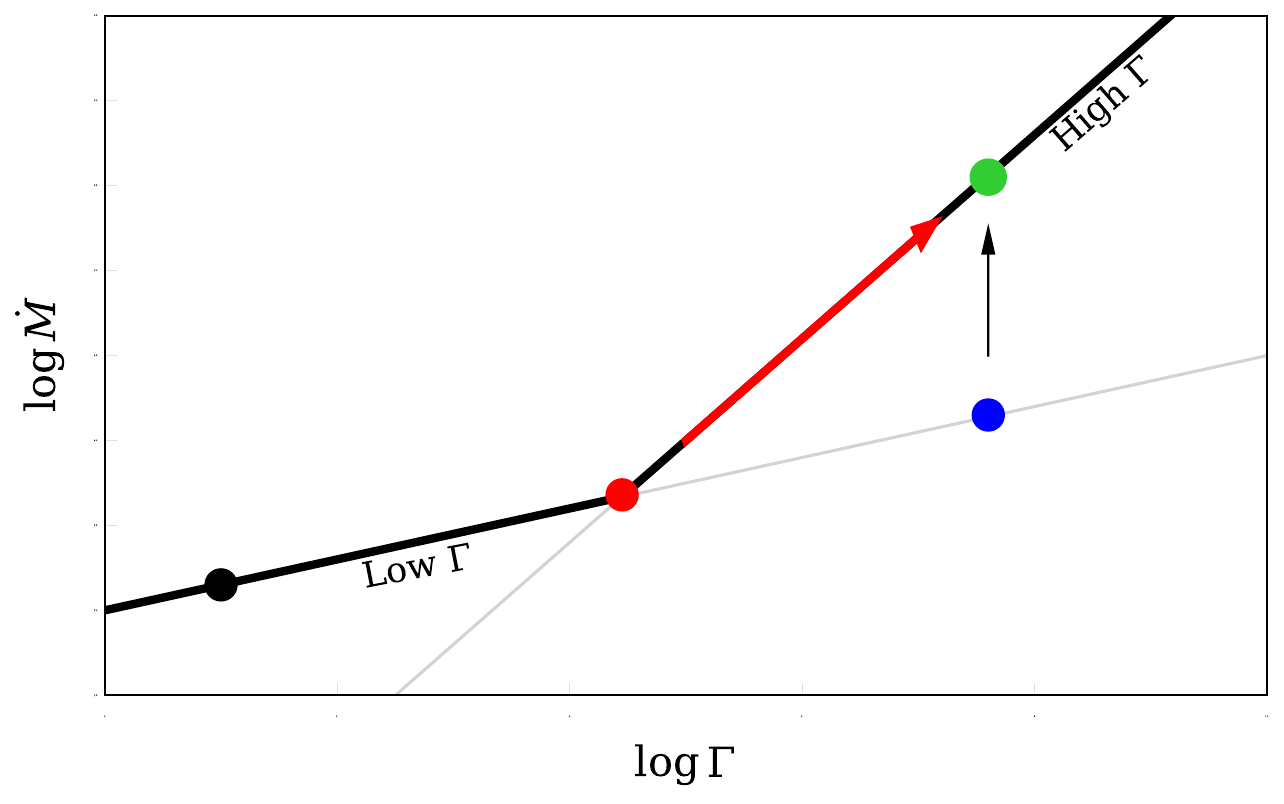}
  \caption{Wind kink implementation. Mass-loss rates follow a CAK-type relation in the low $\Gamma$ regime, but get enhanced when crossing the red dot, where the wind efficiency number $\eta$ crosses unity. The high $\Gamma$ mass-loss rate relation is much steeper (5 instead of 2; \cite{vink2011}).}
  \label{f_carton}
\end{figure}

If one were to have a full data-set listing all   
luminosities for 
Of supergiants, Of/WN transition stars as well as fully-fledged WNh stars, the corresponding transition mass-loss rate $\dot{M}_{\rm trans}$ could be accurately derived by simple determination of the transition luminosity $L_{\rm trans}$ and the terminal velocity $v_{\infty}$ from UV P\,Cygni lines,

\begin{equation}
\dot{M}_{\rm trans} = \frac{L_{\rm trans}}{v_{\infty} c}
\label{eq_transm}
\end{equation}
This transition mass-loss rate can be obtained by very accurate means, 
independent of assumptions regarding the complex physics and implementation of clumping.

\cite{VG12} determined this transition \mdot\ for solar metallicity. They found the spectroscopic transition -- where an absorption line spectrum of an O-type star turns into an emission line at Of/WN -- to occur at $\log(L/L_{\odot}) = 6$, with a mass-loss rate of 
  $\dot{M_{\rm trans}} \simeq 10^{-5} \msunyr$.
The only uncertainty in this transition mass-loss rate analysis involves relatively small errors in 
$v_{\infty}$ (of $\sim$10\%) and $L$ (at most 30\%).
These errors are far smaller than the systematic uncertainties associated with the derivation of empirical mass-loss 
rates resulting from the effects of clumping and porosity.

\section{Empirical mass-loss rates}

In an ideal world we would be in a good position to compare the theoretical mass-loss rates derived from
radiative line driving to empirical mass-loss rates from large homogeneous data-sets.
Unfortunately this is yet not feasible, as empirical analyses have so far involved limited data-sets. Also the analyses have generally been performed using smooth winds, but even when applying clumped-wind prescriptions, they have usually simply treated the clumping in an optically thin "micro-clumping approach" when performing atmospheric spectral analyses. 
It is generally assumed that the mass is confined to high-density clumps that follow a simply $\beta$ velocity law and a void inter-clump medium. Moreover the ionisation balance is generally computed simply from the high-density clumps via a $\rho^2$ scaling of opacities.

\begin{figure}
  \includegraphics[width=\columnwidth]{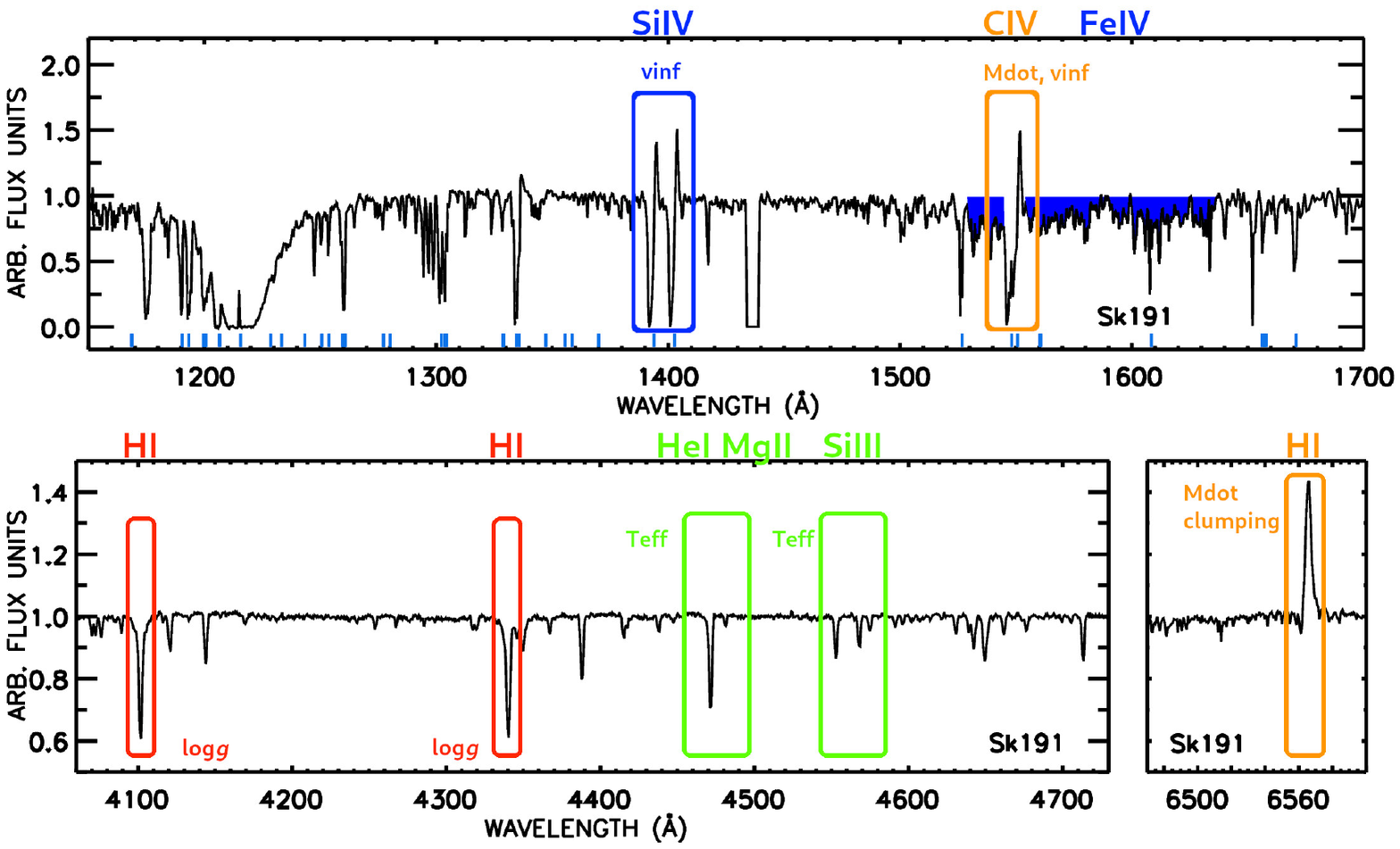}
  \caption{UV (ULLYSES) and optical (XshootU) spectroscopy of the early B supergiant Sk 191 in the SMC, highlighting
key photospheric (red and green) and wind (blue, orange) diagnostics. See \cite{XshootU}.}
  \label{f_xshootu}
\end{figure}

In order to make progress we first need to understand the origin of wind structures. These might involve both sub-surface regions associated with the Fe opacity bump \citep{Cant09,jiang15,moens22} or the LDI \citep{OCR}. The second step would be to obtain the correct physical description of the wind clumping as a function of radius. As a third and final step we may utilise such a parametrisation inside non-LTE spectral atmosphere codes such as Fastwind \citep{Puls20}, PoWR \citep{sander20}, or CMFGEN \citep{Hillier98} and perform combined UV/optical multi-wavelength spectral analyses. 

Large observational samples are currently being gathered by the Hubble Space Telescope ULLYSES project in the UV and the Very Large Telescope X-Shooter instrument in the optical range, as part of the XShootU Project. An example of such a combined UV$+$optical spectrum, and its diagnostic potential is presented in Fig.\,\ref{f_xshootu}. While the stellar parameters such as $T_{\rm eff}$ and $\log{g}$ can be determined from optical absorption lines, for the derivation of the wind terminal velocity, the amount of wind clumping, and the overall mass-loss rate, various P\,Cygni type and emission lines in both the UV and optical range need to be considered simultaneously.

\section{Current Uncertainties}\label{chap1:sec7}%

Let us summarise our knowledge and uncertainties in wind mass loss, and how these uncertainties could affect stellar evolution, by contrast to the alternative uncertainties associated with interior mixing (core-boundary mixing, rotational mixing, etc.).

\begin{BoxTypeA}[chap1:box1]{Mass-loss rate uncertainties in various Mass Regimes}

\begin{itemize}
%\end{BoxTypeA}    
%\section*{Box 1 hd}
%Stated another way, this concept requires that the mass into and out of an infinitesimal box must be equal to the change of mass in the box. Such a volume is sketched,
%\subsection*{Box 2 hd}
%The symbols $\delta x$, $\delta y$, and $\delta z$ represent the
\item{$M_{\rm ZAMS} < 20\msun$}: Here stars are in the  {\it weak-wind regime} \citep{martins05} and it has yet to be established if the root cause is theoretical or diagnostic in nature. As stellar winds are weak in this mass regime, even an uncertainty of 1-2 orders of magnitude should not have a strong effect on stellar mass evolution. Uncertainties in interior mixing are thus expected to dominate in this regime.\\

\item{20$\msun < M_{\rm ZAMS} < 60\msun$}
which we consider to be canonical massive stars. Here the uncertainties in mass-loss rates are substantial ($\sim$3 \citep{krticka17,sundqvist19,vink22}, and as both mixing and mass loss play a role in the evolution of these stars. Disentangling the separate effects of winds and mixing is absolutely crucial here.\\

\item{$M_{\rm ZAMS} > 60\msun$}: In this upper mass range the effect of winds completely dominate the stellar evolution in comparison to mixing. While there could be some uncertainty in \mdot\ at 60\msun, the values around the transition mass-loss point -- which occurs at about 80-100\msun\ for Galactic $Z$ -- are the most accurate (by as little as 30\%).\\

\end{itemize}

\end{BoxTypeA}

\section{Massive star winds at later evolutionary phases}

So far we have focused on the winds of O-type stars, VMS, and classical WR stars. However there are additional evolutionary phases, characterised by the LBV phase, as well as the yellow and red supergiant phases. The driving physics of these phases is still very uncertain at the current time. 

For instance, we are aware of cases -- such as the extreme LBV $\eta$ Car -- where up to 10 solar masses of material was expelled over a period of just a few decades. 
It is hence pivotal to understand the physics of the LBV phenomenon. 
Prior to jumping to the most extreme cases like $\eta$ Car, there is a need to understand the common S Doradus mass-loss cycles, necessary in order to get a realistic impression of the impact of the LBV phenomenon in more general terms.
The terms LBV, episodic and eruptive mass loss, superwind and super-Eddington wind have also been used in SN literature, but with different meanings.

Let us clarify some relevant terminology.
In some cases, gigantic mass loss such as that of $\eta$ Car could be both episodic and eruptive. 
However, generally just because mass loss is episodic this would not automatically imply that it has an eruptive nature. S\,Dor variations for instance exhibit episodic mass loss but their behaviour appears to be well explained by (quasi)stationary winds, without the need to invoke eruptions.

\subsection{Super-Eddington winds}

During quiescent phases LBVs may lose significant amounts of mass via stationary winds, but for objects such as $\eta$ Car more extreme mass loss needs to be considered \citep{Smith14}. 
The famous Homunculus nebula may have been created from
a super-Eddington wind with a mass-loss rate that is three orders of magnitude higher than one could be expected from an ordinary line-driven wind.

Continuum-driven winds in super-Eddington stars could reach mass-loss rates that are so huge that they may approach 
the photon tiring limit where $\mdot_{\rm tir} = L/(GM/R)$. This could result in a stagnating 
outflow that would lead to porous structures.
Alternatively, structures may arise from instabilities relating to subsurface Fe opacity peaks when stars naturally approach the Eddington limit.

The equation of motion can be approximated as: 

\begin{equation}
v\Bigl(1-\frac{v_{\rm s}^2}{v^2}\Bigr)\frac{{\rm d}v}{{\rm d}r} \simeq
g_{\rm grav}(r) + a_{\rm rad}(r) =-\frac{GM}{r^2}(1-\Gamma(r)) .
\end{equation}
At the sonic point, $r_{\rm s}$: $v=v_{\rm s}$, and  
$a_{\rm rad} = -g_{\rm grav}$ implying $\Gamma(r_{\rm s})=1$.

We require $\Gamma$ to cross unity at $r_{\rm s}$, as it needs to fulfil both the condition (i) that it is less than unity near the surface, i.e. $\Gamma(r)$ must be $<1$ below the sonic point, and (ii) that $\Gamma(r)$ rises to values larger than unity in the supersonic region, in order to achieve a stellar wind solution.
This implies an increasing opacity$\frac{{\rm d}\bar \kappa}{{\rm d}r}|_s>0$. 
Without sub-sonic opacity reduction due to porosity, one would not have the situation that $\Gamma$ crosses unity at the sonic point, and the entire atmosphere would be Super-Eddington with $\Gamma(r) >1$, and the star would be blown to pieces.

{\it If} the sub-sonic opacity could be reduced, e.g. due to the formation of a porous medium \citep{shaviv2000} -- which is quite likely near the Eddington limit -- a continuum-driven wind solution becomes viable. 
The reason is that outward travelling photons could avoid the optically thick clumps, lowering $a_{\rm rad}$ 
below the sonic point, and the effective Eddington parameter might drop below unity. 
Further out in the wind, the clumps become optically thinner -- as a result of expansion -- and the porosity effect thus less relevant, and $\Gamma$ could be larger than 1. 
In other words, a wind 
solution with $\Gamma$ crossing unity is feasible, even for 
stars that are formally above the Eddington limit.

\cite{Owocki04} developed a CAK-like model in terms of an effective opacity concept and porosity length. They showed that  
that the mass-loss rate might become substantial when the porosity length achieves a size of order of the 
pressure scale height.  They showed 
that even the huge mass-loss eruption of $\eta$ Car might in principle be explicable by radiative driving, though alternatives might be viable as well. 

\subsection{Cool star winds}

While there are successful radiation-driven wind models off dust 
for the driving of AGB winds \citep{hofner}, such a model does sofar not exist for RSG winds \citep{decin}. 
The general expectation is that there is some way in which to increase
the density scale-height, e.g. by radial pulsations \citep{Wilson84}, and that
the dust opacity at larger radius from the photosphere will drive the wind
to its terminal velocity \citep{Loon05}. The question on the physical mechanism that
sets the mass-loss rate for RSGs in the inner wind remains an active area of research. The physical ingredients may involve convection, turbulence, and shocks \citep{Kee21,Fuller24}.

\section{Outlook} 

In line with its use by the AGB community, the term superwind describes the cumulative
effect of increasing wind mass-loss rates when stars become more luminous. These winds could be radiation-driven and of increasing strength when they approach the Eddington limit \citep{GV16}. 
RSGs could also be subjected to additional wind initiation mechanism, involving convection, turbulence, shocks, pulsations, etc. 

We finally mention an empirical mass-loss "kink" feature uncovered by \cite{Yang23}, and we highlight its similarity to radiation-driven winds at the optically thin/thick transition point for hot stars. 
\cite{VS23} motivated a mass-loss prescription that depends on the Eddington
factor $\Gamma$ (including both a steep $L$-dependence and an inverse steep $M$-dependence), which could play a relevant role during the stellar evolution of massive red and other cool supergiants. We should nonetheless be wary of the fact that the general challenge of identifying the correct physical force(s) initiating massive RSG winds, thereby correctly interpreting Eq.\,\ref{eq_eom}, as well as the issue of the absolute $\dot{M}$ determinations \citep{Beasor20}, still persists in this cool-star regime of the HR diagram.

\begin{ack}[Acknowledgments] 
I I would like to thank Ethan Winch, Erin Higgins, and Gautham Sabhahit for their help with some of the figures and  their general contributions to the Armagh Mdot group. I would like to extend my gratitude to the XShootU collaboration. Fabian Scheider is thanked for providing useful editorial comments.
\end{ack}

\seealso{\cite{puls08,owocki15,vink22,Smith14,decin}}

\bibliographystyle{Harvard}
\bibliography{stellar-winds}

\end{document}